\newcommand{\ve}{\mathbf}

\newcommand{\veo}[1]{ \hat{\mathbf #1}}

\newcommand{\ket}[1]{\left| #1\right>}
\newcommand{\bra}[1]{\left< #1\right|}
\newcommand{\braket}[2]{ \left< #1 | #2 \right > }
\newcommand{\melement}[3]{ \left< #1 | #2 | #3 \right> }
\newcommand{\welement}[2]{(#1|#2)_{Q_z}}
\newcommand{\eps}{\varepsilon}
\newcommand{\floor}[1]{\lfloor #1 \rfloor}
\newcommand{\hc}{\mathrm{h.c.}}
\newcommand{\nsum}{\sum_{n_{1234}}}
\newcommand{\diss}{\hat{\mathcal{D}}}

\RequirePackage{fix-cm}
\documentclass[twocolumn]{svjour3}          

\smartqed  
\usepackage{graphicx}
\usepackage{amsmath}
\usepackage{color}
\usepackage{amssymb}

\journalname{Journal of Computational Electronics}
\begin{document}


\title{Partially coherent electron transport in terahertz quantum cascade lasers based on a Markovian master equation for the density matrix}

\titlerunning{Density-matrix master equation for terahertz QCL simulation} 

\author{O. Jonasson  \and F. Karimi \and
        I. Knezevic }


\institute{O. Jonasson$^{1}$, F. Karimi$^{2}$ and I. Knezevic$^{3}$ \at
University of Wisconsin -- Madison \\
Madison, Wisconsin 53706, USA\\
              $^{1}$\email{ojonasson@wisc.edu} \\
              $^{2}$\email{karimi2@wisc.edu} \\
              $^{3}$\email{irena.knezevic@wisc.edu}
}

\date{Received: date / Accepted: date}

\maketitle

\begin{abstract}
We derive a Markovian master equation for the single-electron density matrix, applicable to quantum cascade lasers (QCLs). The equation conserves the positivity of the density matrix, includes off-diagonal elements (coherences) as well as in-plane dynamics, and accounts for electron scattering with phonons and impurities. We use the model to simulate a terahertz-frequency QCL, and compare the results with both experiment and simulation via nonequilibrium Green's functions (NEGF). We obtain very good agreement with both experiment and NEGF when the QCL is biased for optimal lasing. For the considered device, we show that the magnitude of coherences can be a significant fraction of the diagonal matrix elements, which demonstrates their importance when describing THz QCLs. We show that the in-plane energy distribution can deviate far from a heated Maxwellian distribution, which suggests that the assumption of thermalized subbands in simplified density-matrix models is inadequate. We also show that the current density and subband occupations relax towards their steady-state values on very different time scales. 
\end{abstract}
\section{Introduction}
Quantum cascade lasers (QCLs) are semiconductor heterostructures that operate based on quantum confinement and tunneling. Population inversion between quasi-bound lasing states is achieved through precise engineering of material composition and layer widths~\cite{faist_science_1994}. Numerical simulations play an important role in the design of QCLs~\cite{dupont_PRB_2010,jirauschek_APR_2014}. For this purpose, a range of theoretical models have been employed, including semiclassical ~\cite{iotti2001,callebaut_APL_2004,gao_JAP_2007,shi_JAP_2014} and quantum-transport techniques based on the density matrix formalism~\cite{willenberg_PRB_2003,kumar_PRB_2009,weber_PRB_2009,dupont_PRB_2010,terazzi_NJP_2010}  or nonequilibrium Green's functions (NEGF)~\cite{lee_PRB_2002}. Semiclassical approaches are appealing due to their low computational requirements. They go beyond the effective-mass approximation~\cite{gao_JAP_2007} and can explore phenomena such as nonequilibrium phonons~\cite{shi_JAP_2014}. However, semiclassical models can provide an inadequate descriptions to QCLs working in the THz range, where the role of coherence cannot be ignored~\cite{callebaut_JAP_2005,kumar_PRB_2009}.

In order to maximize the performance of THz QCLs, optimization methods such as genetic algorithms have been used, where the simulation converges on a layer structure that maximizes the gain of the device~\cite{dupont_JAP_2012}. These simulations require repeated calculations of device performance for a large number of parameters, so computational efficiency plays an important role. This fact makes density-matrix-based approaches advantageous over the relatively high computational burden of NEGF~\cite{lindskog_APL_2014}. However, common density-matrix-based approaches have two significant drawbacks. One is a  common assumption of thermalized subbands, where the electron temperature is either an input parameter~\cite{kumar_PRB_2009,lindskog_APL_2014} or determined using an energy-balance method~\cite{harrison_JAP_2002}. This approximation may not be warranted, because QCLs operate far from equilibrium, so the in-plane energy distribution can (as will be shown later in this work) can deviate far from a heated thermal distribution (Maxwellian or Fermi-Dirac), making electron temperature an ill-defined quantity. The second drawback is phenomenological treatment of dephasing
\cite{kumar_PRB_2009,lindskog_APL_2014,terazzi_NJP_2010,callebaut_JAP_2005}.

In this work, we propose a computationally efficient density-matrix model based on a rigorously derived Markovian master equation. The Markovian master equation  conserves the positivity of the density matrix, includes off-diagonal matrix elements as well as full in-plane dynamics and time-dependence, and accounts for the relevant scattering mechanisms with phonons and impurities. We apply the model on a terahertz QCL proposed in Ref.~\cite{dupont_JAP_2012}. With the QCL biased for lasing, we obtain very good agreement with experiment,  as well as theoretical results based on NEGF. We show that the magnitude of off-diagonal elements of the density matrix (coherences) can be a significant fraction of the diagonal values, demonstrating the importance of including coherence when describing THz QCLs. We show that significant electron heating takes place, where the in-plane energy distribution of subbands deviates far from a thermal distribution, with  each subband having a unique energy dependence. Lastly, we provide time-resolved results, giving insight into the response of the device to a suddently applied bias, revealing the different time scales involved.

This paper is organized into $5$ sections and an appendix. In Sec.~\ref{sec:derivation}, we derive a Markovian master equation for the single-electron density matrix that is applicable to electron transport in QCLs. In Sec.~\ref{sec:numerical}, we describe the the numerical solution method. Results for a THz QCL are given in Sec.~\ref{sec:results}, along with comparison to NEGF and experiment. Section~\ref{sec:conclusion} contains concluding remarks.

\section{Derivation of the master equation}
\label{sec:derivation}

In the following, we will denote three-dimensional (3D) vectors with uppercase letters and two-dimensional (2D) vectors as lowercase letters. For example, $\ve Q=(Q_x,Q_y,Q_z)$ and $\ve k=(k_x,k_y)$, where transport is in the \textit{z}-direction (cross-plane) and translational invariance in the \textit{x}-\textit{y} plane (in-plane) direction is assumed. $\ve Q+\ve k$ should be understood as $(Q_x+k_x,Q_y+k_y,Q_z)$.

The total Hamiltonian of an open electronic system, describing the behavior of electrons interacting with a dissipative phonon bath can be written as
\begin{equation}
 \hat H=\hat H_0 + \hat H_\mathrm{e-ph} + \hat H_\mathrm{ph}.
\end{equation}
$\hat H_0$ is the unperturbed Hamiltonian of electrons, including the kinetic and potential electronic terms, and $\hat H_\mathrm{ph}$ denotes the Hamiltonian of the free phonon bath. The interaction Hamiltonian between electrons and phonons is included in $\hat {H}_{\text{e-ph}}$.

We use a Fr\"{o}lich-type Hamiltonian to describe the interaction of a single electron with a phonon bath:~\cite{frolich_PRSA_1937}
\begin{align}
 \hat H_\mathrm{e-ph}=\frac{1}{(2\pi)^3} \sum_{g} \int d^3Q\,  \mathcal{M}_{g}(\mathbf{Q}) (b_{g,\mathbf{Q}}e^{i\mathbf{Q \cdot \hat{R}}}-b^\dagger_{g,\mathbf{Q}}e^{-i\mathbf{Q \cdot \hat{R}}}) \ .
\end{align}
Here, $b^\dagger_{g,\mathbf{Q}}$ ($b_{g,\mathbf{Q}}$) is the phonon creation (annihilation) operator for a phonon in branch $g$ with wave vector $\ve Q$ and $\mathcal M_g(\ve Q)$ is the associated scattering matrix element. Note that we have assumed the phonon wave vectors are closely spaced to warrant integration over $\ve Q$. The equation of motion for the statistical operator ($\hat{\rho}$) in the interaction picture is
\begin{align}
\nonumber
\frac{d}{dt} \tilde{\hat{\rho}}(t)=&-\frac{i}{\hbar}[\hat {\tilde{H}}_\text{e-ph}(t),\tilde{\hat{\rho}}(t)],\\ \label{eom1}
\tilde{\hat{\rho}}(t)=&\tilde{\hat{\rho}}(0)-\frac{i}{\hbar}\int_0^t[\hat {\tilde{H}}_\text{e-ph}(t'),\tilde{\hat{\rho}}(t')]dt' \ .
\end{align}
The tilde symbol denotes that the operators are in the interaction picture, i.e.,  $\hat{\tilde{O}}(t)=e^{\frac{i}{\hbar}(\hat{H}_0+\hat{H}_{\text{ph}})t}\hat{O}e^{-\frac{i}{\hbar}(\hat{H}_0+\hat{H}_\text{ph})t}$.
We assume the interaction of the electron and phonons only negligibly affects the density matrix of the phonon reservoir (\emph{Born approximation}), thus the density matrix of the total system may be represented as a tensor product $ \tilde{\hat{\rho}}(t)=\tilde{\hat{\rho}}_e(t)\otimes \tilde{\hat{\rho}}_\text{ph}$ \cite{breuer2002theory,knezevic_JCEL_2013}. We also assume the interaction strength is sufficiently high to treat the system as memoryless (\emph{Markov approximation}), i.e., the evolution of the density matrix only depend on its present state. Now, we put the integral form  in Eq. \eqref{eom1} in the right hand side of the differential form, then we apply the Born and Markov approximations, and finally we take the trace over the phonon reservoir. Then, the equation of motion reads
\begin{align}
\label{eom3}
&\frac{d}{dt}\tilde{\hat{\rho}}_e(t)=-\frac{i}{\hbar}\text{tr}_\text{ph}\left\{[\hat {\tilde{H}}_\text{e-ph}(t),\tilde{\hat{\rho}}_e(0)\otimes \tilde{\hat{\rho}}_\text{ph}]\right\}\\ \nonumber
&-\frac{1}{\hbar^2}\int_0^\infty ds~\text{tr}_\text{ph}\left\{[\hat {\tilde{H}}_\text{e-ph}(t),[\hat {\tilde{H}}_\text{e-ph}(t-s),\tilde{\hat{\rho}}_e(t)\otimes \tilde{\hat{\rho}}_\text{ph}]]\right\}.
\end{align}
In order to remove the temporal dependence of the interaction Hamiltonian, we switch back to the Schr\"{o}dinger picture, and use
$\text{tr}_\text{ph}\left\{\hat {{H}}_\text{e-ph} {\hat{\rho}}_\text{ph}\right\}=0$, giving
\begin{align}
\label{eom10}
&\frac{d\hat{\rho}_e(t)}{dt}=-\frac{i}{\hbar}[\hat {H}_{\text{0}}, \hat{\rho}_e(t) ] -\frac{1}{\hbar^2}\int_0^\infty ds~ \times \\ \nonumber
&\text{tr}_\text{ph}\left\{[\hat {{H}}_\text{e-ph},[e^{-i(\hat{H}_0+\hat{H}_\text{ph})s/\hbar} \hat {{H}}_\text{e-ph}e^{i(\hat{H}_0+\hat{H}_\text{ph})s/\hbar},{\hat{\rho}}_e(t)\otimes {\hat{\rho}}_\text{ph}]]\right\}.
\end{align}
We will refer to the second term on the right hand side of the above equation as $\diss$, the \textit{dissipation superoperator} or \textit{the dissipator}, acting on the density matrix. The equation of motion for the reduced single-electron density operator $\hat \rho_e$ can then be written as
\begin{align}
  \label{eq:lioscatt}
  \frac{\partial \hat \rho_e}{\partial t}=-\frac{i}{\hbar}[\hat H_0,\hat \rho_e] + \diss(\hat \rho_e) \ ,
\end{align}
where $\diss$ contains the effect of of dissipation due to interactions with phonons (static disorder can also be included in $\diss$ \cite{kohn1957,fischetti1999,Karimi_2016}). By tracing over the phonon degree of freedom in~\eqref{eom10} and expanding the commutators, $\diss$ can be grouped into eight terms, containing four hermitian conjugate pairs. Two terms correspond to emission and two to absorption. In order to keep the equations compact, calculations will only be shown explicitly for the emission terms. Using this simplification we can write
\begin{align}
  \nonumber
  \diss(\hat \rho_e) &= \frac{1}{\hbar^2}\frac{1}{(2\pi)^3}\int d^3Q\int_{0}^{\infty}ds \{ \\ \nonumber
    &- \mathcal W_g^\mathrm{em}(\ve Q)e^{-i E_g s/\hbar}  e^{-i\ve Q\cdot \veo R }e^{-i\hat H_0 s/\hbar}e^{i\ve Q\cdot \veo R}e^{i\hat H_0 s/\hbar}\hat \rho_e  \\ \nonumber
       &+ \mathcal W_g^\mathrm{em}(\ve Q)e^{+iE_g s/\hbar} e^{i\ve Q\cdot \veo R}\hat \rho_e e^{-\hat H_0 s/\hbar}e^{-i\ve Q\cdot \veo R}e^{i\hat H_0 s/\hbar} \\ \label{eq:beforebasis}
       &  + \hc + \mathrm{abs.} \} \ ,
\end{align}
where abs. refers to absorption terms and $\mathcal W_g^\mathrm{em}(\ve Q)=| \mathcal{M}_{g}(\ve Q)|^2 (N_g+1)$, with $E_g$ the phonon energy and $N_g=(e^{E_g/k_BT}-1)^{-1}$ the phonon occupation. The absorption terms can be obtained in the end by flipping the sign of the phonon energy $E_g$ and making the switch $N_g+1\rightarrow N_g$. The two terms in Eq.~\eqref{eq:beforebasis} correspond to out-scattering (first term, negative sign) and in-scattering (second term, positive sign). The appendix gives $\mathcal W_g^\mathrm{em}(\ve Q)$ for various interaction mechanisms.

To proceed, we pick the eigenstates of $\hat H_0$ as a basis. The eigenstates are denoted as $\ket{n,\ve k}=\ket{n}\otimes\ket{\ve k}$, where $n$ labels the discrete set of eigenfunctions with energy $E_n$ in the $z$-direction (subband energies) and $\ve k$ labels the continuous set of free-particle eigenfunctions with energy $E_k=\hbar^2k^2/2m^*$ in the in-plane direction with the effective mass $m^*$. The phase of the basis states is chosen such that $\psi_n(z)=\braket{z}{n}$ are real. With this choice of basis, we have $\braket{n',\ve k'}{n,\ve k}=\delta_{n'n}\delta(\ve k'-\ve k)$. We assume translational invariance in the in-plane direction so both the density matrix and the dissipator are diagonal in $\ve k$
\begin{subequations}
\begin{align}
  \label{eq:densdiag}
  \melement{n',\ve k'}{\hat \rho_e}{n,\ve k}&=  \rho_{n'n}^{E_k}\delta(\ve k'-\ve k) \\
  \melement{n',\ve k'}{\diss}{n,\ve k}&= \mathcal{D}_{n'n}^{E_k}\delta(\ve k'-\ve k) \ ,
\end{align}
\end{subequations}
where the matrix elements of $\hat \rho_e$ and $\hat{\mathcal D}$ are labeled according to their energy $E_k$. In order to make the following derivation more compact we define the the following quantities
\begin{subequations}
\begin{align}
 (n|m)_{Q_z}&= \melement{n}{e^{iQ_z \hat z}}{m} \\
 \Delta_{nm}&=E_n-E_m \\
 E(n,\ve k)&=E_n+E_k \ .
\end{align}
\end{subequations}
In Sec.~\ref{ssec:outscatt}, we simplify the out-scattering term in Eq.~\ref{eq:beforebasis} and do the same for the in-scattering term in Sec.~\ref{ssec:inscatt}. In section~\ref{ssec:perio}, we write the master equation in a form applicable to periodic systems such a QCLs.

\subsection{Out-scattering term}
\label{ssec:outscatt}
We will start with the out-scattering-term, which is the first term in Eq.~\eqref{eq:beforebasis}. By using the completeness relation $4$ times, we can write the dissipator term corresponding to emission due to interaction mechanism $g$ as
\begin{align}
  \nonumber
  \diss_\mathrm{em,g}^\mathrm{out}&=-\frac{1}{\hbar^2(2\pi)^3} \int d^3Q \int_0^\infty ds \nsum \int d^2k_{1234} \times \\ \nonumber
  &\melement{n_1,\ve k_1}{ e^{-i\ve Q\cdot \veo R} e^{-i\hat H_0 s/\hbar} }{n_2,\ve k_2}  e^{-iE_g s/\hbar} \times \\ \nonumber
  &\melement{n_2,\ve k_2}{ e^{+i\ve Q\cdot \veo R} e^{+i\hat H_0 s/\hbar} }{n_3,\ve k_3} \mathcal W_g^\mathrm{em}(\ve Q) \times \\
  &\melement{n_3,\ve k_3}{\hat \rho_e}{n_4,\ve k_4}\ket{n_1,\ve k_1}\bra{n_4,\ve k_4} + \hc \ ,
  \label{eq:afterbasis}
\end{align}
where $\int d^2 k_{1234}$ refers to integration over $\ve k_1$ through $\ve k_4$ and $n_{1234}$ refers to sum over $n_1$ through $n_4$. We can simplify the above expression using
\begin{align}
  \nonumber
  \melement{n,\ve k}{ e^{\pm i\ve Q\cdot \veo R} e^{\pm i\hat H_0 s/\hbar} }{n',\ve k'}&= \\
  (n|n')^*_{Q_z}e^{\pm iE(n',\ve k')s/\hbar}&\delta[\ve k-(\ve k'\pm \ve q)] \ .
  \label{eq:firstelement}
\end{align}
Using Eq.~\eqref{eq:firstelement} and after performing the $\ve k_4$ integration, Eq.~\eqref{eq:afterbasis} becomes
\begin{align}
\nonumber
&\diss_\mathrm{em,g}^\mathrm{out}=-\frac{1}{\hbar^2(2\pi)^3}\int
  \int d^3Q \int_0^\infty ds \nsum d^2k_{123} \times \\ \nonumber
& \mathcal W_g^\mathrm{em}(\ve Q)
  \welement{n_1}{n_2}^*\welement{n_2}{n_3} \times \\ \nonumber
&e^{-i\frac{s}{\hbar} \left( E(n_2,\ve k_2)-E(n_3,\ve k_3)+E_g \right)}
  \rho_{n_3n_4}^{E_{k_3}}\ket{n_1,\ve k_1}\bra{n_4,\ve k_3} \times \\
& \delta[\ve k_1-(\ve k_2-\ve q)]\delta[\ve k_2-(\ve k_3+\ve q)]
  + \hc \ .
\end{align}
After performing the $\ve k_2$ and $\ve k_3$ integration, we get
\begin{align}
\nonumber
&\diss_\mathrm{em,g}^\mathrm{out}=-\frac{1}{\hbar^2(2\pi)^3}
 \int d^3Q \int_0^\infty ds \nsum \int d^2k_1 \times \\ \nonumber
& \mathcal W_g^\mathrm{em}(\ve Q) \welement{n_1}{n_2}^*\welement{n_2}{n_3}
  \rho_{n_3n_4}^{E_{k_1}} \ket{n_1,\ve k_1}\bra{n_4,\ve k_1} \times \\
& e^{-i\frac{s}{\hbar} \left( E(n_2,\ve k_1+\ve q)-E(n_3,\ve k_1)+E_g \right)   } + \hc \ .
\end{align}
In order to perform the $s$ integration, we use
\begin{align}
  \int_0^\infty e^{-i\Delta \frac{s}{\hbar}} ds = \pi\hbar \delta(\Delta)  -i\hbar \mathcal P\frac{1}{\Delta} \ ,
\end{align}
where $\mathcal P$ denotes the Cauchy principal value, which leads to a small correction to energies (Lamb shift) \cite{breuer2002theory}. Ignoring the principa-value term and shifting the integration variable $\ve Q\rightarrow \ve Q-\ve k_1$, we get
\begin{align}
\nonumber
&\diss_\mathrm{em,g}^\mathrm{out}=-\frac{\pi}{\hbar(2\pi)^3}
  \int d^3Q \nsum \int d^2k_1 \times \\ \nonumber
&\mathcal W_g^\mathrm{em}(\ve Q-\ve k_1)
  \welement{n_1}{n_2}^*\welement{n_2}{n_3}\rho_{n_3n_4}^{E_{k_1}} \times \\
&\delta[ \Delta_{n_2n_3}+E_q-E_{k_1} + E_g ]
  \ket{n_1,\ve k_1}\bra{n_4,\ve k_1} + \hc \ .
\end{align}
Sandwiching both sides by $\melement{N,\ve k}{...}{M,\ve k'}$, integrating over $\ve k_1$ and $\ve k'$and renaming the sum variables $n_2\rightarrow m$, $n_3\rightarrow m$ gives
\begin{align}
\nonumber
&[\diss_\mathrm{em,g}^\mathrm{out}]_{NM}^{E_k} = -\sum_{n,m} \rho_{nM}^{E_{k}}
  \frac{\pi}{\hbar(2\pi)^3} \int d^3Q \ \mathcal W_g^\mathrm{em}(\ve Q-\ve k)\times  \\
&\delta[ \Delta_{mn}+E_q-E_k + E_g ] \welement{N}{m}^*\welement{m}{n} + \hc \ ,
\label{eq:beforegamma}
\end{align}
where in this context, h.c. means "switch $N$ and $M$ and perform complex conjugation". We can write Eq.~\eqref{eq:beforegamma} more compactly as
\begin{align}
[\diss_\mathrm{em,g}^\mathrm{out}]_{NM}^{E_k}
  = -\sum_{n} \rho_{nM}^{E_k} \Gamma^\mathrm{out}_\mathrm{em,g}(N,n,E_k) +  \hc \ ,
\end{align}
with
\begin{align}
\nonumber
\Gamma^\mathrm{out}_\mathrm{em,g}&(N,n,E_k)=\frac{\pi}{\hbar(2\pi)^3}
  \sum_m \int d^3Q \ \mathcal W_g^\mathrm{em}(\ve Q,E_k) \times \\
&\delta[ \Delta_{mn}+E_q-E_k + E_g ] \welement{N}{m}^*\welement{m}{n} \ ,
\label{eq:temp1}
\end{align}
where we have written $\mathcal W_g^\mathrm{em}(\ve Q-\ve k)=\mathcal W_g^\mathrm{em}(\ve Q,E_k)$ because the coordinate system for the $\ve q$ integration can be chosen relative to $\ve k$ so $\mathcal W_g^\mathrm{em}$ only depends on the magnitude of $\ve k$. Note that $\Gamma^\mathrm{out}_\mathrm{em,g}$ has the units of inverse time and is real. These terms will be referred to as \textit{rates} from now on. The rates do not depend on the density matrix, so they can be precalculated and stored.

The $Q_z$ integration in Eq.~\eqref{eq:temp1} involves inner products, such as $(m|n)_{Q_z}$, and has to performed numerically. However, the in-plane integration can be done analytically, so it is useful to rewrite Eq.~\eqref{eq:temp1} as
\begin{align}
\nonumber
&\Gamma^\mathrm{out}_\mathrm{em,g}(N,n,E_k)=\frac{\pi}{\hbar(2\pi)^3}
  \sum_m \int_{-\infty}^\infty d Q_z (N|m)^*_{Q_z} \times \\
&(m|n)_{Q_z} \int d^2 q \delta[\Delta_{nm}+E_k-E_g-E_q]\mathcal W_g^\mathrm{em}(\ve Q,E_k) \ .
\end{align}
The real (imaginary) part of the integrand is even (odd), so we can limit the range of integration to positive $Q_z$. Switching to polar coordinates $d^2q\rightarrow qdq \ d\theta$, making a change of variables $E_q=\hbar^2q^2/2m^*$ and performing the $E_q$ integration gives
\begin{align}
\nonumber
\Gamma^\mathrm{out}_\mathrm{em,g}(N,n,E_k) &=\frac{m^*}{2\pi\hbar^3}
  \sum_m \theta(\Delta_{nm}-E_g+E_k) \times \\ \nonumber
&\int_0^\infty d Q_z \mathop{Re} \left[(N|m)^*_{Q_z}(m|n)_{Q_z}\right] \times \\
&\mathcal G_g^\mathrm{em}(E_k,Q_z,\Delta_{nm}-E_g+E_k) \ ,
\label{eq:inplaneint}
\end{align}
where $\theta$ the Heaviside function and
\begin{align}
  \label{eq:G_def1}
  \mathcal G_g^\mathrm{em}(E_k,Q_z,E_q)=\frac{1}{2\pi} \int_{0}^{2\pi}d\theta\mathcal W_g^\mathrm{em}(\ve Q,E_k) \ ,
\end{align}
where $\mathcal W(\ve Q,E_k)$ can always be written in terms of $Q_z$, $E_k$, $E_q=\hbar^2q^2/2m^*$ and the polar angle $\theta$ of $\ve q$. The explicit form of the function $\mathcal G_g^\mathrm{em}$ depends on the scattering mechanism $g$, and is calculated in appendix~\ref{app:Gterms} for acoustic phonons, nonpolar optical phonons, polar optical phonons (POP), and ionized impurities.

\subsection{In-scattering term}
\label{ssec:inscatt}
By using the completeness relation four times, the in-scattering term in Eq.~\eqref{eq:beforebasis} becomes
\begin{align}
\nonumber
&\diss_\mathrm{em,g}^\mathrm{in}=\frac{1}{\hbar^2(2\pi)^3}
  \int d^3Q \int_0^{\infty} ds \nsum \int d^2k_{1234} \times \\ \nonumber
& \mathcal W_g^\mathrm{em}(\ve Q) e^{iE_gs/\hbar}
  \melement{n_1,\ve k_1}{e^{i\ve Q\cdot \veo r}}{n_2,\ve k_2} \times \\ \nonumber
& \melement{n_3,\ve k_3}{ e^{-i\hat H_0s/\hbar}e^{-i\ve Q\cdot \veo r}
  e^{i\hat H_0s/\hbar}}{n_4,\ve k_4} \times \\
& \melement{n_2,\ve k_2}{\hat \rho_e}{n_3,\ve k_3}  \ket{n_1,\ve k_1}\bra{n_4,\ve k_4} + \hc \ .
\end{align}
Using Eq.~\eqref{eq:densdiag} and
\begin{align}
\nonumber
&\melement{n_3,\ve k_3}{ e^{-i\hat H_0s/\hbar}e^{-i\ve Q\cdot \veo r} e^{i\hat H_0s/\hbar} }{n_4,\ve k_4}= \\
&e^{-i\frac{s}{\hbar}( E(n_3,\ve k_3)-E(n_4,\ve k_4)  )}(n_3|n_4)^*_{Q_z}\delta[\ve k_4-(\ve k_3+\ve q)] ,
\end{align}
gives (after performing the $\ve k_4$ integration)
\begin{align}
\nonumber
\diss_\mathrm{em,g}^\mathrm{in}&=\frac{1}{\hbar^2(2\pi)^3}
  \int d^3Q \int_0^{\infty} ds \nsum \int d^2k_{123} \times \\ \nonumber
& \mathcal W_g^\mathrm{em}(\ve Q)(n_1|n_2)_{Q_z}(n_3|n_4)^*_{Q_z} \times \\ \nonumber
& \rho_{n_2n_3}^{E_{k_2}}e^{-i\frac{s}{\hbar}(E(n_3,\ve k_3)-E(n_4,\ve k_3+\ve q)-E_g)}
  \delta[\ve k_2-\ve k_3] \times  \\
& \ket{n_1,\ve k_1}\bra{n_4,\ve k_3+\ve q} + \hc \ .
\end{align}
Performing the $\ve k_3$ and $\ve k_2$ integrations gives
\begin{align}
\nonumber
&\diss_\mathrm{em,g}^\mathrm{in}=\frac{1}{\hbar^2(2\pi)^3}
  \int d^3Q \int_0^{\infty} ds \nsum \times \\ \nonumber
&\int d^2k_1 \mathcal W_g^\mathrm{em}(\ve Q)(n_1|n_2)_{Q_z}(n_3|n_4)^*_{Q_z}
  \rho_{n_2n_3}^{E_{ |\ve k_1-\ve q| }} \times \\
& e^{-i\frac{s}{\hbar}(E(n_3,\ve k_1-\ve q)-E(n_4,\ve k_1)-E_g)} \ket{n_1,\ve k_1}\bra{n_4,\ve k_1} + \hc \ .
\end{align}
Changing the $\ve Q$ integration variable $\ve Q\rightarrow \ve -\ve Q+\ve k_1$ and performing the $s$ integration (ignoring the principal value) gives
\begin{align}
\nonumber
&\diss_\mathrm{em,g}^\mathrm{in}=\frac{\pi}{\hbar(2\pi)^3} \int d^3Q  \nsum
  \int d^2k_1 \mathcal W_g^\mathrm{em}(\ve Q-\ve k_1) \times \\ \nonumber
& (n_1|n_2)_{Q_z}(n_3|n_4)^*_{Q_z}\rho_{n_2n_3}^{E_q}\delta[\Delta_{n_3n_4}-E_{k_1}-E_g+E_q] \times \\
& \ket{n_1,\ve k_1}\bra{n_4,\ve k_1} + \hc \ ,
\end{align}
where we have used $\mathcal W_g^\mathrm{em}(-\ve Q+\ve k_1)=\mathcal W_g^\mathrm{em}(\ve Q-\ve k_1)$. Sandwiching both sides by $\melement{N,\ve k}{...}{M,\ve k'}$, integrating over $\ve k'$ and $\ve k_1$ and renaming the dummy variables $n_2\rightarrow n$ and $n_3\rightarrow m$ gives
\begin{align}
\nonumber
[\diss_\mathrm{em,g}^\mathrm{in}]_{NM}^{E_k} =&  \sum_{n,m} \rho_{nm}^{E_k+E_g+\Delta_{Mm}}
  \frac{\pi}{\hbar(2\pi)^3} \int d^3Q \times \\ \nonumber
& \mathcal W_g^\mathrm{em}(\ve Q-\ve k)(N|n)_{Q_z}(m|M)^*_{Q_z} \times \\
& \delta[\Delta_{Mm}+E_{k}+E_g-E_q] + \hc \ .
\end{align}
After doing the in-plane integration over $\ve q$, we get
\begin{align}
  \nonumber
  [\diss_\mathrm{em,g}^\mathrm{in}]_{NM}^{E_k} &= \sum_{n,m} \rho_{nm}^{E_k+E_g+\Delta_{Mm}}
  \Gamma_\mathrm{em,g}^\mathrm{in}(N,M,n,m,E_k) \\
  &+ \hc \ ,
\end{align}
where we have defined the in-scattering analog of Eq.~\eqref{eq:inplaneint}
\begin{align}
\nonumber
\Gamma_\mathrm{em,g}^\mathrm{in}(N,M,n,m,&E_k)
  =\frac{m^*}{2\pi\hbar^3}\theta[ \Delta_{Mm}+E_g+E_k ] \times \\ \nonumber
&\int_0^\infty dQ_z \mathop{Re}\left[ (N|n)_{Q_z}(m|M)^*_{Q_z} \right] \times \\
&\mathcal G_g^\mathrm{em}(E_k,Q_z,\Delta_{Mm}+E_g+E_k) \ ,
\label{eq:gamma_in}
\end{align}
with $\mathcal G_g^\mathrm{em}$ defined in Eq.~\eqref{eq:G_def1}.

\subsection{Application to periodic systems}
\label{ssec:perio}
The Markovian master equation (MME) for the density matrix elements can be written by summing over all different scattering mechanisms $g$;
\begin{align}
\frac{\partial \rho_{NM}^{E_k}}{\partial t}= -i\frac{\Delta_{NM}}{\hbar}\rho_{NM}^{E_k}+\mathcal{D}_{NM}^{E_k} \ ,
\end{align}
with
\begin{align}
\nonumber
\mathcal{D}_{NM}^{E_k}= &-\sum_{g,n} \Gamma_\mathrm{em,g}^\mathrm{out}(N,n,E_k) \rho_{nM}^{E_k}\\ \nonumber
& + \sum_{n,m,g}\Gamma^\mathrm{in}_\mathrm{em,g}(N,M,n,m,E_k)\rho_{nm}^{E_k+E_g+\Delta_{Mm}} \\
& +\hc+ \mathrm{abs.} \ ,
\label{eq:MME_1}
\end{align}
where abs. refers to absorption terms and the in and out-scattering rates are defined in Eqs.~\eqref{eq:inplaneint} and \eqref{eq:gamma_in} respectively. The $\mathcal G_g^\mathrm{em}$ functions are calculated in appendix~\ref{app:Gterms} for various scattering mechanisms.

The form of the MME in Eq.~\eqref{eq:MME_1} is not well suited for periodic systems such as QCLs. It is more conenient to work with relative indices
\begin{align}
  \label{eq:f_def}
  f_{N,M}^{E_k}\equiv \rho_{N,N+M}^{E_k} \ .
\end{align}
Using relative indices, it is easy to take advantage of periodicity, where
\begin{align}
  f_{N,M}^{E_k}=f_{N\pm N_s,M}^{E_k} \ .
\end{align}
The range $N\in [1,N_s]$ is the number of eigenstates $N_s$ in a single period. The choice of which period to consider is arbitrary but in this work we choose the center period corresponding to the range $z\in[-L_p/2,L_p/2]$, where $L_p$ is the period length. A state is considered to be in the center period if  $|\melement{n}{\hat z}{n}|\leq L_p/2$, i.e., if the state's center of mass is in the center period. The elements with $M=0$ give the diagonals of the density matrix and $M\neq0$ gives the coherence a distance of $M$ from the diagonal. The $M$ indice runs from $-\infty$ to $+\infty$ so a truncation needs to be performed in order to do numerical calculations. Truncation of $M$ will be discussed in Sec.~\ref{ssec:coherence_cut}.

Inserting Eq.~\eqref{eq:f_def} into Eq.~\eqref{eq:MME_1} gives
\begin{align}
\nonumber
\frac{\partial f_{N,M}^{E_k}}{\partial t}= -i\frac{\Delta_{N,N+M}}{\hbar} f_{N,M}^{E_k}
-\sum_{n,g} \Gamma^\mathrm{out,em,g}_{NMnE_k} f_{N,n}^{E_k}   \\
+ \sum_{n,m,g}\Gamma^\mathrm{in,em,g}_{NMnmE_k} f_{N+n,M+m-n}^{E_k+E_g+\Delta_{Mm}}+\hc+ \mathrm{abs.}
\label{eq:MMEPC}
\end{align}
with
\begin{subequations}
\begin{align}
\nonumber
&\Gamma^\mathrm{out,em}_{NMnE_k}=\frac{m^*}{2\pi\hbar^3}\sum_{m}
  \theta[\Delta_{N\!+n,N\!+M\!+m}\!-E_g\!+E_k] \int_0^\infty \times \\ \nonumber
& d Q_z \mathop{Re} \left[(N\!+\!M|N\!+M\!+m)^*_{Q_z}(N\!+M\!+m|N\!+n)_{Q_z}\right] \times  \\
& \mathcal G_g^\mathrm{em}(E_k,Q_z,\Delta_{N+n,N+M+m}-E_g+E_k) \ ,
\label{eq:goc}
\end{align}
and
\begin{align}
\nonumber
&\Gamma^\mathrm{in,em,g}_{NMnmE_k} =
  \frac{m^*}{2\pi\hbar^3}\theta[ \Delta_{N\!+M,N\!+M\!+m}\!+E_g\!+E_k ] \times  \\ \nonumber
&\int_0^\infty\! dQ_z \mathop{Re}\left[ (N|N\!+n)_{Q_z}(N\!+M\!+m|N\!+M)^*_{Q_z} \right] \times  \\
&\mathcal G_g^\mathrm{em}(E_k,Q_z,\Delta_{N+M,N+M+m}+E_g+E_k) \ .
\label{eq:gic}
\end{align}
\end{subequations}
Note that in Eqs.~\eqref{eq:goc} and \eqref{eq:gic}, the  dummy indices $n$ and $m$ have been shifted in such a way that terms with large $n$ or $m$ are small. Equation~\eqref{eq:MMEPC} (with accompanying Eqs. \eqref{eq:goc} and \eqref{eq:gic}) is the main result in  this work. In the next section, we will discuss numerical solution methods for Eq. ~\eqref{eq:MMEPC}. For evaluation of $\mathcal G_g^\mathrm{em}$ for various interaction mechanisms, we refer the reader to the appendix.

\section{Numerical method}
\label{sec:numerical}

The central quantity is the density matrix $f_{NM}^{E_k}$ which is stored for $N\in[1,N_s]$, $M\in[-N_c,N_c]$, $E_k\in[0,E_\mathrm{max}]$. Here, $N_c$ is an integer that quantifies how far apart in energy the states can be to still have appreciable off-diagonal density-matrix terms (coherences); we refer to $N_c$ as the coherene cutoff. $E_\mathrm{max}$ is the in-plane kinetic-energy cutoff. The energies are discretized into $N_E$ evenly spaced values, such that the density matrix array has dimensions $N_s\times(2N_c+1)\times N_E$.

The basic idea is to start with a chosen initial state and numerically time-step Eq.~\eqref{eq:MMEPC}, until a steady state is reached. For the time stepping, we use an asynchronous leapfrog method, which is a robust second order, two-step, explicit method for the integration of the Liouville equation~\cite{mutze2013}. This choice of the time-stepping method allows us to use a rather large time step of $1$~fs, which is about $10$ times larger than an Euler time-stepping scheme would allow.

Note that the sums in the MME ~\eqref{eq:MMEPC} run over matrix elements and energies outside the fundamental period (e.g., $N>N_s$ or $N<1$), which are calculated using the modulo operation
\begin{subequations}
\begin{align}
  f_{N,M}^{E_k}&=f_{N',M}^{E_k} \\
  E_N&=E_{N'}+\frac{N-N'}{N_s}E_0 \\
  N'&=\mathrm{mod}(N-1,N_s)+1 \ ,
\end{align}
\end{subequations}
with $\mathrm{mod}(n,N_s) = n - N_s \floor{n/N_s}$ and $E_0$ the potential energy drop over a single period (intrinsic function MOD in Matlab and MODULUS in gfortran). The MME~\eqref{eq:MMEPC} also contains terms for which $|M|>N_c$, where we assume $f_{NM}^{E_k}=0$.

\subsection{Coherence cutoff and performance}
\label{ssec:coherence_cut}
Equation~\eqref{eq:MMEPC} contains an infinite sum that represents coupling between eigenstates over infinitely long distances. However, it is easy to see that terms with small $|n|$ and $|m|$ are dominant. For example, the in-scattering term contains terms on the form $(N|N\!+n)_{Q_z}$ and $(N\!+M\!+m|N\!+M)^*_{Q_z}$, which are small for large $|n|$ and $|m|$ respectively due to the low spatial overlap of states that are highly seperated in energy. For the same reason, out-scattering terms with high $|n|$ or $|m|$ are small, too. In this work, we truncate the sum by only including terms with $|n|,|m|\leq N_c$. We note that the the numerical method could be improved by only summing over a subset of $n,m\in[-N_c,N_c]$, that contains the biggest rates.

From Eq.~\eqref{eq:MMEPC}, we see that the in-scattering term is the bottleneck in the time-evolution of the density matrix. The computational complexity for the time-evolution is  $\mathcal O(N_sN_c^3N_EN_g)$, and therefore depends most strongly on the coherence cutoff, $N_c$. The computational complexity only depends linearly on the number of eigenstates $N_s$, which opens the possibility to study multiple periods of QCLs and investigate effects of electric field domain formation~\cite{wacker_PR_2002}, which has a negative effect of QCL performance. The minimum coherence cutoff needed for convergence is highly system-dependent. In this work, a modest value of $N_c=5$ proved to be sufficient for convergence in current and occupations. Other parameters used in this work are $N_E=101$, $N_s=5$, and $N_g=4$. The number of time steps is $10^5$, with a time step of $1$~fs, resulting in $100$~ps of simulated time. Using these parameters, typical simulations times for a single value of the electric field were about $45$ minutes on an Intel Core i7-2600 (gfortran complier, running on a single core). As mentioned before, the simulation time could be reduced significantly by only summing over a chosen small subset of $n,m\in[-N_c,N_c]$ in Eq.~\eqref{eq:MMEPC}.

\subsection{Initial state}

We choose an initial state corresponding to thermal equilibrium. Assuming Boltzmann statistic, the density matrix factors into in-plane and cross-plane terms and we can write
\begin{align}
  \left. f_{NM}^{E_k}\right|_\mathrm{eq}=C_{NM}e^{-E_k/k_BT} \ .
\end{align}
To calculate the expansion coefficients $C_{NM}$, we first solve for the Bloch states $\phi_{s,q}(z)$ ($s$ labels the band and $q\in \left[-\pi/L_p,\pi/L_p\right]$ labels the wave vector in the Brillouin zone  associated with the structure's period $L_p$) by diagonalizing the Hamiltonian in~\eqref{eq:schrod} with $V_B(z)=0$, using a basis of plane waves. We can then calculate the cross-plane equilibrium density matrix using
\begin{align}
  \rho_\mathrm{eq}(z_1,z_2)= \sum_s \int_{-\pi/L_p}^{\pi/L_p}\phi_{s,q}(z_1)\phi^*_{s,q}(z_2) e^{-E_{s,q}/k_BT}dq \ .
\end{align}
Using the above result, we can calculate the expansion coefficients
\begin{align}
  C_{NM}= \int d z_1 d z_2 \psi_{N}(z_1) \psi_{N+M}(z_2) \rho_\mathrm{eq}(z_1,z_2) \ .
\end{align}
This choice of initial condition works well with an electric field that is turned on instantaneously at time $t=0^+$; this  is limiting case of an abruptly turned-on bias. If only the steady state is sought, all terms with $M\neq0$ can be artificially set equal to zero in the initial density matrix; this initial condition avoids high-amplitude coherent oscillations during the transient and leads to a faster numerical convergence towards the steady state.

\subsection{Bandstructure calculation}
\label{sec:ee}

Upon the application of bias, we assume the field and the associated linear potential drop are established instantaneously, but that the density matrix and charge distribution take a while to respond and do so adiabatically.

We treat the eigenstates under an applied bias as bound states, even though, strictly speaking, the states are better described as resonances with some energy spread~\cite{moiseyev_PR_1998}. The bound-state approximation is good if the energy spread is much smaller than other characteristic energies, and if the dynamics are mostly limited to the subspace of resonance states. For more discussion on the validity of this approximation, see Ref.~\cite{jirauschek_APR_2014}.

The eigenstates and subband energies are obtained from the Schr\"{o}dinger equation
\begin{align}
\nonumber
&\left( -\frac{\hbar^2}{2} \frac{d}{dz}\frac{1}{m(z)}\frac{d}{d z} + V_\mathrm{SL}(z) + V_\mathrm{B}(z) + V_\mathrm{H}(z,t) \right) \psi_n(z,t)\\
&=E_n(t) \psi_n(z,t) \ ,
\label{eq:schrod}
\end{align}
where $m^*(z)$ is a position-dependent effective mass, $V_\mathrm{SL}$ is the superlattice potential (wells and barriers), $V_\mathrm{B}$ the linear potential drop due to an applied bias, and $V_\mathrm{H}$ the mean-field Hartree potential, which is obtained by solving Poisson's equation. The Hartree potential $V_H$ depends on the electron density and is therefore time-dependent. However, its time evolution is weak due to low doping and is typically very slow, so we can assume that the adiabatic approximation holds and the concept of eigenstates and energies is well defined during the transient. In  writing Eq.~\eqref{eq:schrod}, we have neglected coupling of of the eigenfunctions with the in-plane motion, which is a standard assumption when describing QCLs and other superlattices~\cite{wacker_PR_2002,jirauschek_APR_2014}.

To calculate the eigenfunctions under bias, which we assume are fairly well localized, we use a basis of Hermite functions (eigenfunctions of the harmonic oscillator) and diagonalize the Hamiltonian in Eq.~\eqref{eq:schrod}.
Recalculating the eigenfunctions and computing the new rates,  Eqs.~\eqref{eq:goc} and \eqref{eq:gic}, as the system evolves is a computationally expensive procedure, taking about $10$ to $50$ times longer than a single time step. However, it does not need to be done in every time step due to the slow temporal and spatial variation of $V_\mathrm{H}$. In order to recalculate the eigenfunctions only when needed, we calculate
\begin{align}
  \delta=\max_m \left| \int dz |\psi_m(z,t_\ell)|^2(V_\mathrm{H}(z,t_\ell)-V_\mathrm{H}(z,t_{\ell-1})) \right| \ ,
\end{align}
where $t_\ell$ is the time at the current time step $\ell$. The quantity $\delta$ is the magnitude of the maximal first-order energy correction to the eigenstates. If $\delta$ is above a certain threshold energy, we recalculate the eigenfunctions and the corresponding rates. If the threshold is not met, we do not update the wavefunctions nor the Hartree potential. The procedure of calculating $\delta$ is very cheap in terms of computational resources and does not noticably affect performance. Typically, the wavefunctions are recalculated frequently during the initial transient and much less frequently near the steady state.  In the present work, we used a threshold energy of $0.1$~meV. This choice of theshold energy typically leads to $\sim 100$ recalculations of eigenfunctions while the total number of time steps is on the order of $10^5$.

When recalculating the eigenfunctions, an issue arises when numbering the updated states and choosing their phase. The time evolution of the eigenfunctions must be adiabatic so the same phase must be chosen for each state when the eigenfunctions are recalculated. Since the eigenfunctions are chosen to be real, there are only two choices of phase. A very simple assigning method is to calculate
\begin{align}
  \alpha_{nm} = \int dz \psi_n(z,t_\ell) \psi_m(z,t_{\ell-1}) \ .
\end{align}
States with the highest overlap $|\alpha_{nm}|\simeq 1$ are "matched" according to $n\rightarrow m$, which ensures the proper numbering of the new states and $\psi_n(z,t_{\ell})\rightarrow \mathrm{sign}(\alpha_{nm})\psi_n(z,t_{\ell})$ takes care of the choice of phase.

\subsection{Low-energy thermalization}

\label{sec:thermalization}
Out of the included scattering mechanisms (POP, acoustic phonons, and ionized impurities) only POP scattering is inelastic. However, the POP energy is typically larger than $k_BT$ and this lack of a low-energy inelastic scattering mechanism leads to numerical difficulties, where in-plane energy distributions can vary abruptly (this problem is often encountered in density-matrix models;  see, for example, Ref.~\cite{weber_PRB_2009}). A detailed inclusion of electron-electron interaction would solve this issue, where arbitrarily low energy can be exchanged between electrons. The small energy exchanges involved in electron-electron interaction also plays a crucial role in thermalization  within a subband. However, electron-electron interaction is a two-body interaction that is not straightforward to include in a single-electron picture.  For this reason, in the present work we will include low energy thermalization (LET) in a simplified manner, by adding a scattering mechanism with an energy equal to the minimal in-plane energy spacing $\Delta E$ in the simulation. The purpose of this extra scattering mechanism is to help smoothen the in-plane energy distribution. We treated the LET as an additional POP-like scattering term, with energy exchange equal to $\Delta E$ and an effective strength denoted by the dimensionless quantity $\alpha$. The matrix element is
\begin{align}
  \label{eq:fict}
  |\mathcal M_\mathrm{LET}(\ve Q)|^2= \alpha
  \frac{e^2\Delta E}{2\eps_0} \left( \frac{1}{\eps_r^\infty} - \frac{1}{\eps_r} \right)
  \frac{Q^2}{(Q^2+Q_D^2)^2} \ ,
\end{align}
where $Q_D^2=ne^2/(\eps k_B T)$ is the Debye wave vector. The reason for the choice a POP-like matrix element is its preference for small-$\ve Q$ scattering, just like electron-electron interaction. The role of the LET term is mainly to smoothen of the in-plane energy distribution. As we will show later, the results are not very sensitive to the value of $\alpha$.

\section{Results}
\label{sec:results}
\begin{figure}
\centering
\includegraphics[width=3.3in]{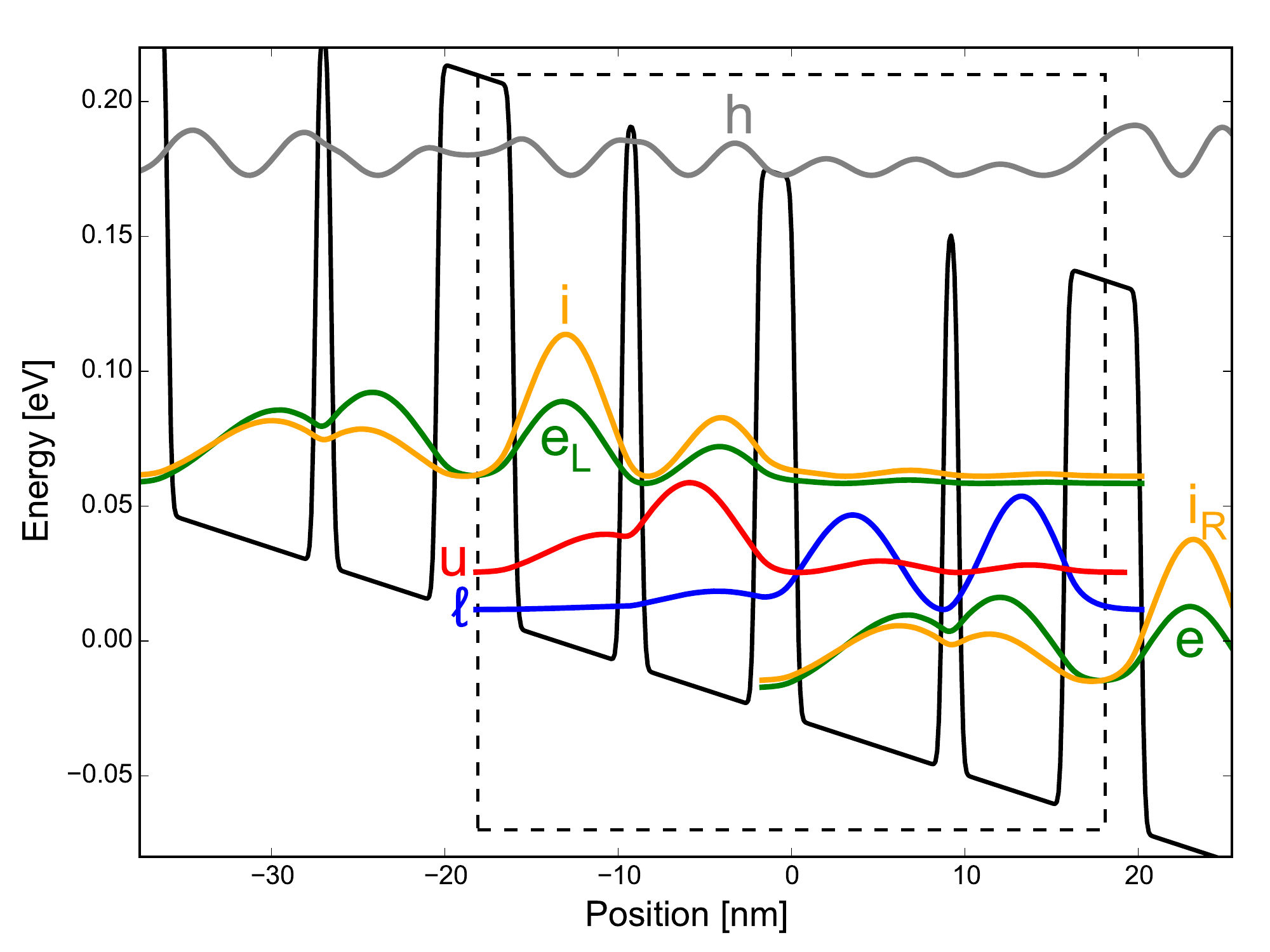}
\caption{Conduction band edge (solid black line) and probability densities for the upper lasing state (u), lower lasing state ($\ell$), injector state (i), and extractor state (e). Also shown is the extractor state (e$_L$) for the previous stage to the left, the injector (i$_R$) state for the next stage to the right, and a high-energy state (h). The high-energy state was included in numerical calculations, however, it had a small  occupation and a negligible effect on physical observables. The dashed rectangle represents a single stage with the layer structure (from the left) $\mathbf{44}/62.5/\mathbf{10.9}/66.5/\mathbf{22.8}/84.8/\mathbf{9.1}/61$~\AA, with barriers in bold font. The thickest barrier (injector barrier) is doped with Si such that the average electron density is is $8.98\times10^{15}$~cm$^{-3}$. Due to low doping, the potential drop is approximately uniform. }
\label{bandstructure}       
\end{figure}

To demonstrate the validity of our model, we simulated a THz QCL proposed in Ref.~\cite{dupont_JAP_2012}. The authors used a phonon-assisted injection and extraction design based on a GaAs/Al$_{0.25}$Ga$_{0.75}$As material system and achieved lasing at $3.2$~THz, up to a heatsink temperature of $138$~K. We chose this specific device because both experimental and theoretical results are readily available for comparison. Figure~\ref{bandstructure} shows the conduction band profile and most important eigenfunctions of the considered device at the design electric field of $21$~kV/cm. We will split this section into two parts, starting with steady state results in section~\ref{sec:steady} and time resolved results in section~\ref{sec:time}.

\subsection{Steady-state results}
\label{sec:steady}
\begin{figure}
\centering
\includegraphics[width=3.3in]{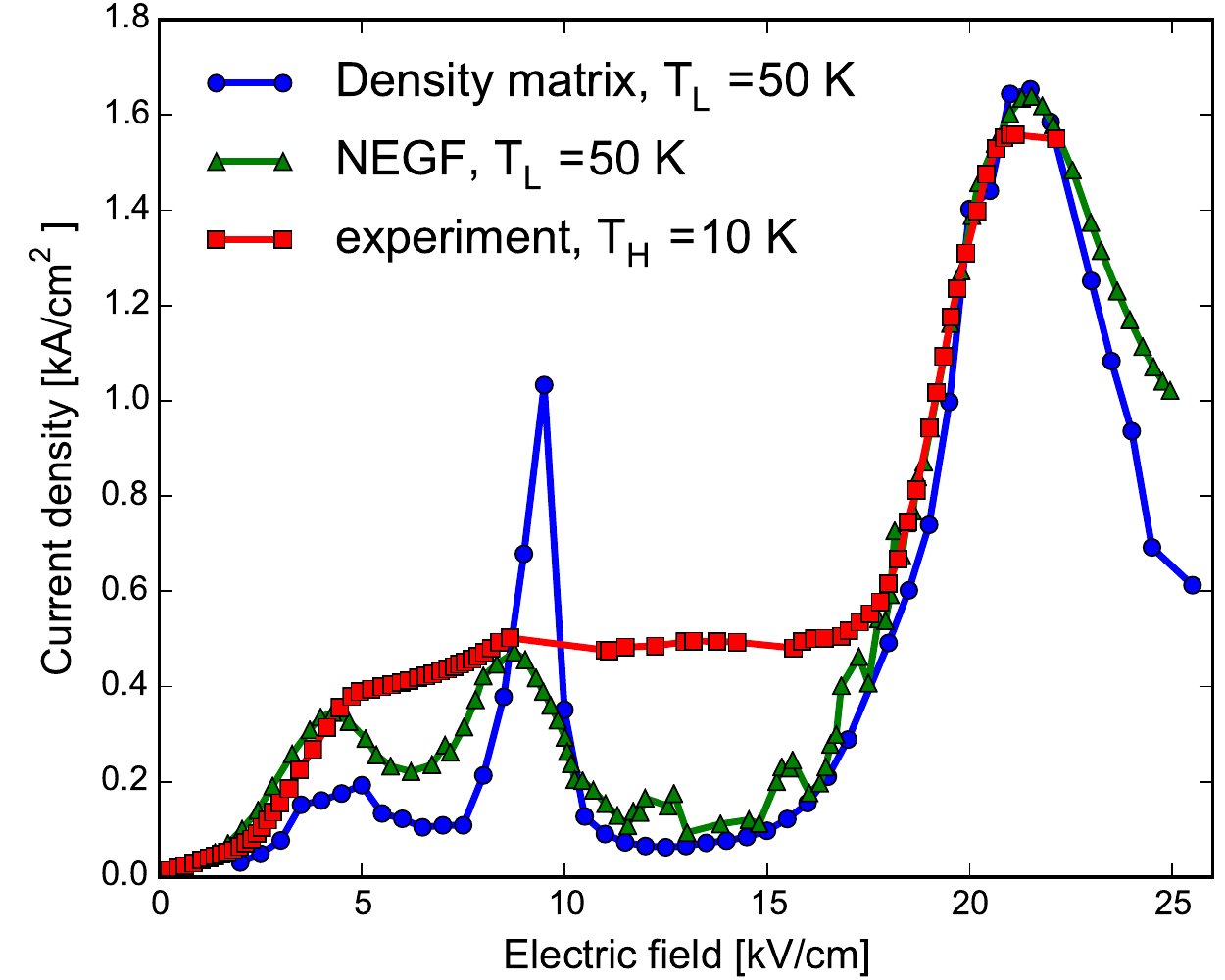}
\caption{Current density vs electric field for density matrix results (blue circles) and NEGF (green triangles) for a lattice temperature $T_L=50$~K. Also shown are experimental results (red squares) for a heat-sink temperature $T_H=10$~K. Experimental and NEGF results are both from Ref.~\cite{dupont_JAP_2012}.}
\label{exp_comparison}       
\end{figure}

\begin{figure}
\centering
\includegraphics[width=3.3in]{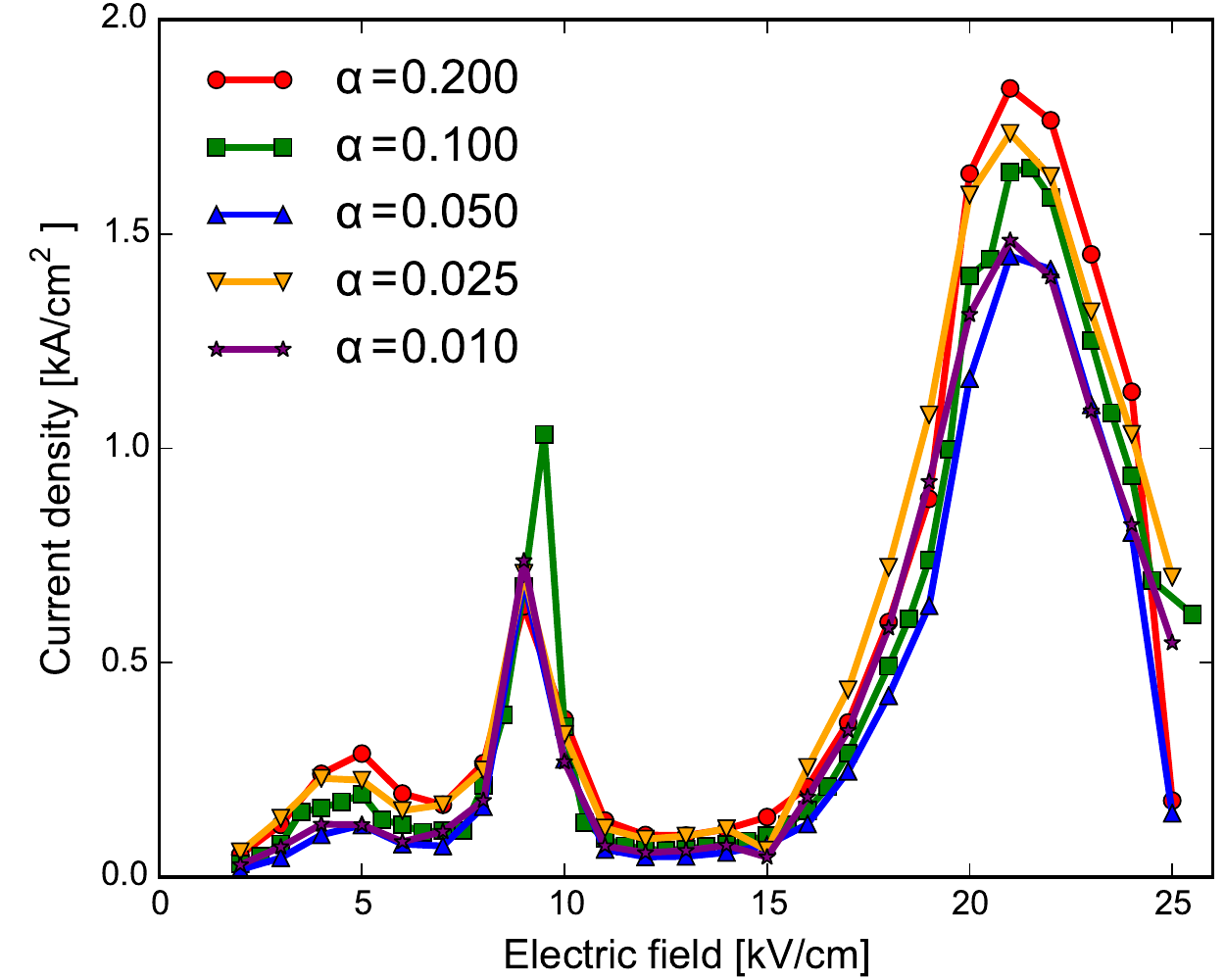}
\caption{Current density vs electric field for different values of the strength parameter $\alpha$. The best agreement with experimental data is for $\alpha=0.1$ (green squares). The results at high fields are not sensitive to the strength parameter, while low-bias results are. The higher peak at $9$~kV/cm for the $\alpha=0.1$ data is a result of a finer electric field mesh for that data set.}
\label{smoothing_effect}       
\end{figure}

Figure~\ref{exp_comparison} shows a steady-state current density vs electric field, as well as comparison with experiment and theoretical results based on NEGF~\cite{dupont_JAP_2012}. The experimental data is for a heat-sink temperature of $T_H=10$~K. The actual lattice temperature $T_L$ is expected to be higher \cite{shi_JAP_2014}. Both the density matrix and NEGF results are for a lattice temperature of $50$~K. We included interactions with polar optical phonons, acoustic phonons (using elastic and equipartition approximations), and ionized impurities. In addition we included a LET scattering mechanism discussed in section~\ref{sec:thermalization} with a strength parameter of $\alpha=0.1$. This choice of $\alpha$ gave the best agreement with experiment. However, results around the design electric field did not depend strongly on $\alpha$, as can be seen in Fig.~\ref{smoothing_effect}. From Fig.~\ref{exp_comparison} we see a very good agreement with experiment and NEGF around the design electric field of $21$~kV/cm. For electric fields lower than $17$~kV/cm, neither NEGF or our density matrix results accurately reproduce experimental results. However, our density matrix results and the NEGF results both show a double-peak behavior. The difference between the density matrix results and NEGF can be attributed to collisional broadening (not captured with density matrix approaches) and our calculation not including interface-roughness scattering.

\begin{figure}
\centering
\includegraphics[width=3.3in]{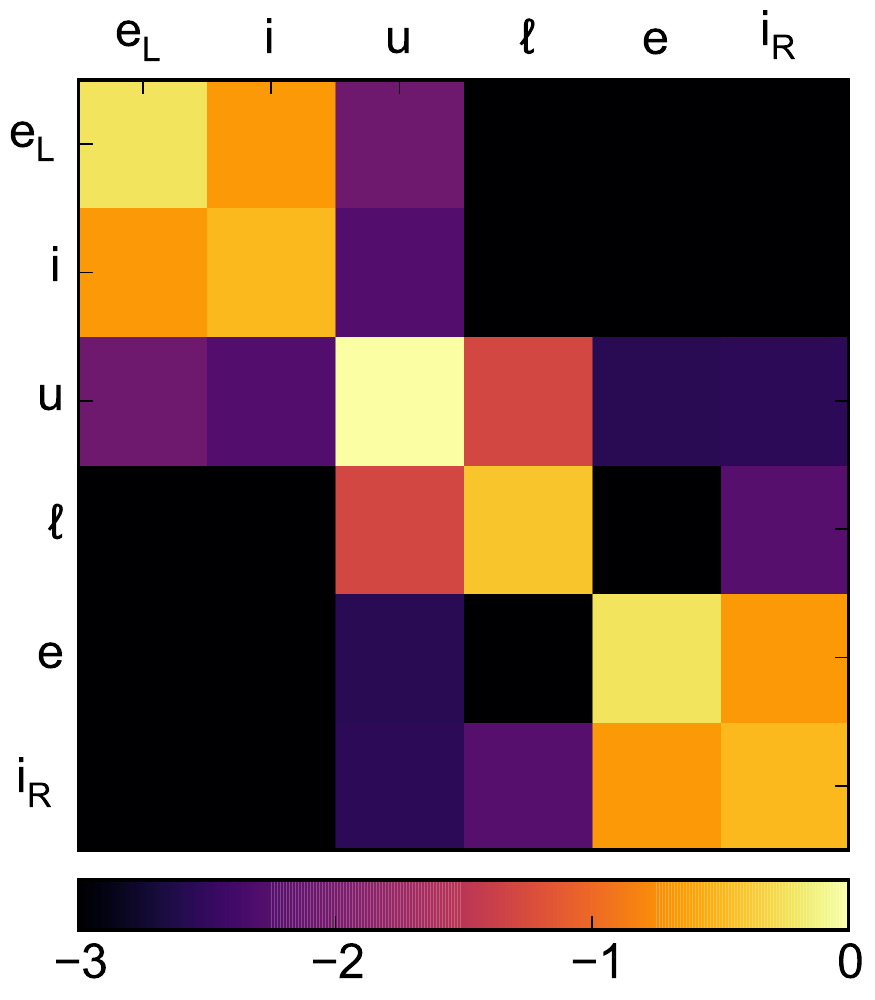}
\caption{$\mathrm{log}_{10}(|\rho_{N\!M}|)$, where $\rho_{N\!M}$ are density matrix elements after interation over parallel energy. Normalization is chosen so that the highest occupation is one (upper lasing level). Results are for an electric field of $21$~kV/cm and a lattice temperature of $50$~K. Coherences and occupatons are given for for all combinations of states shown in Fig.~\ref{bandstructure}, except for the high energy h state, which had very small coherences and occupation. The highest occupations are the upper lasing level ($1.0$), extractor state ($0.60$), lower lasing level ($0.39$) and injector state ($0.33$). The biggest coherences are between the extractor state e and injector state i$_R$ ($0.21$), and between the upper and lower lasing levels ($0.05$). Other coherences were smaller than $0.01$. }
\label{coherence_integrated}       
\end{figure}

In order to visualize the occupations and coherences of all combinations of the states, it is instructive to plot density matrix elements after integrating out the parallel energy
\begin{align}
 \rho_{N\!M}= \int \rho_{N\!M}^{E_k} d E_k \ .
\end{align}
Figure~\ref{coherence_integrated} shows a plot of $\mathrm{log}_{10}(|\rho_{N\!M}|)$, with occupations and coherences of all combinations of the states shown in Fig~\ref{bandstructure}, except for the high-energy (h) state, which had negligible occupation and coherences. Normalization is chosen such that the largest matrix element is 1 (the occupation of the upper lasing level). From the figure, we can see that the magnitude of the coherences can be quite large. For example, the largest  coherence is between the e$_L$ extractor state and the i injector state, with a magnitude of about $0.21$; this is a  significant fraction of the largest  diagonal element and demonstrates the importance of including coherences in calculations. The second largest coherence is between the upper and lowing lasing states, with a magnitude of $0.05$. Other coherences are smaller than $1\%$ of the largest diagonal element and all coherences more than 4 places off the diagonal were smaller than $10^{-3}$, justifying our coherence cutoff of $N_c=5$.

\begin{figure}
\centering
\includegraphics[width=3.3in]{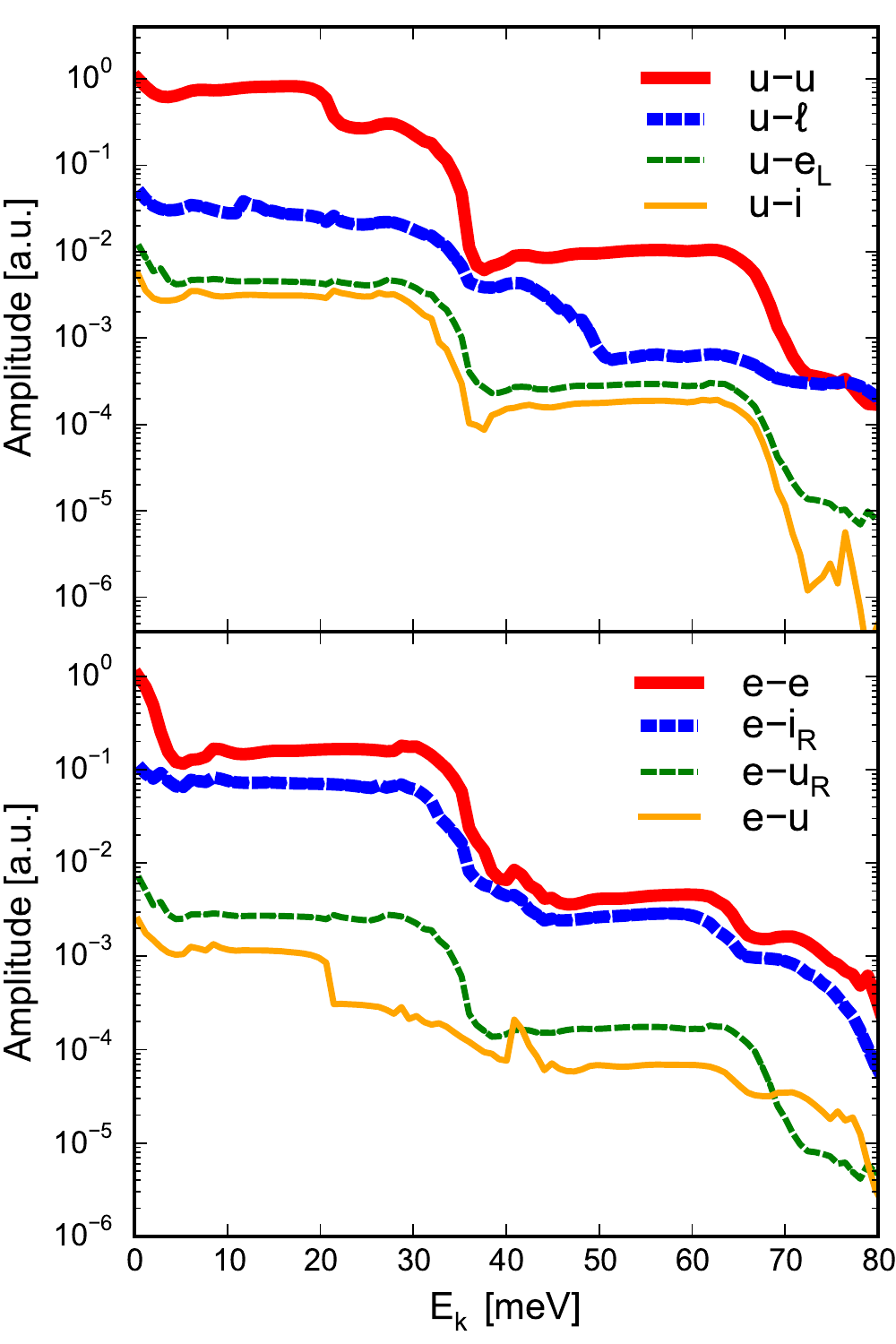}
\caption{The magnitude of the matrix element $|\rho_{N\!M}^{E_k}|$ as a function of in-plane energy, for $N$ corresponding to the upper lasing level (top panel) and extractor state (bottom panel). These two states were chosen because they have the greatest occupation. Note that u$_R$ corresponds to the upper lasing level in the next period to the right (not shown in Fig.~\ref{bandstructure}), which is equal to the coherence between e$_L$ and u owing to periodicity.}
\label{coherence_inplane}
\end{figure}

The magnitude of the matrix elements $\rho_{N\!M}$ give information about the importance of including off-diagonal matrix elements in QCL simulations. However these matrix elements do not give us information about the dependence on in-plane energy. In order to visualize the in-plane dependence, Fig.~\ref{coherence_inplane} shows plots of $\rho_{N\!M}^{E_k}$ as a function of the in-plane energy for multiple pairs of $N$ and $M$. In the top (bottom) panel, $N=u$ ($N=e$) is fixed and $M$ varied, showing the three largest  coherences, as well as the diagonal term. The figure shows the energy dependence of the two largest coherences mentioned earlier (e-i$_R$ and u-l), along with the second and third largest coherences for each state. We see that most off-diagonal elements are more than two orders of magnitude smaller than the diagonal terms. Both the diagonal elements and the coherences have an in-plane distribution that deviates strongly from a Maxwellian distribution, with a sharp drop around $35$~meV due to enhanced POP emission. This result suggests that simplified density-matrix approaches, where a Maxwellian in-plane distribution is assumed, are not justified for the considered system.

\subsection{Time-resolved results}
\label{sec:time}
\begin{figure}
\centering
\includegraphics[width=3.3in]{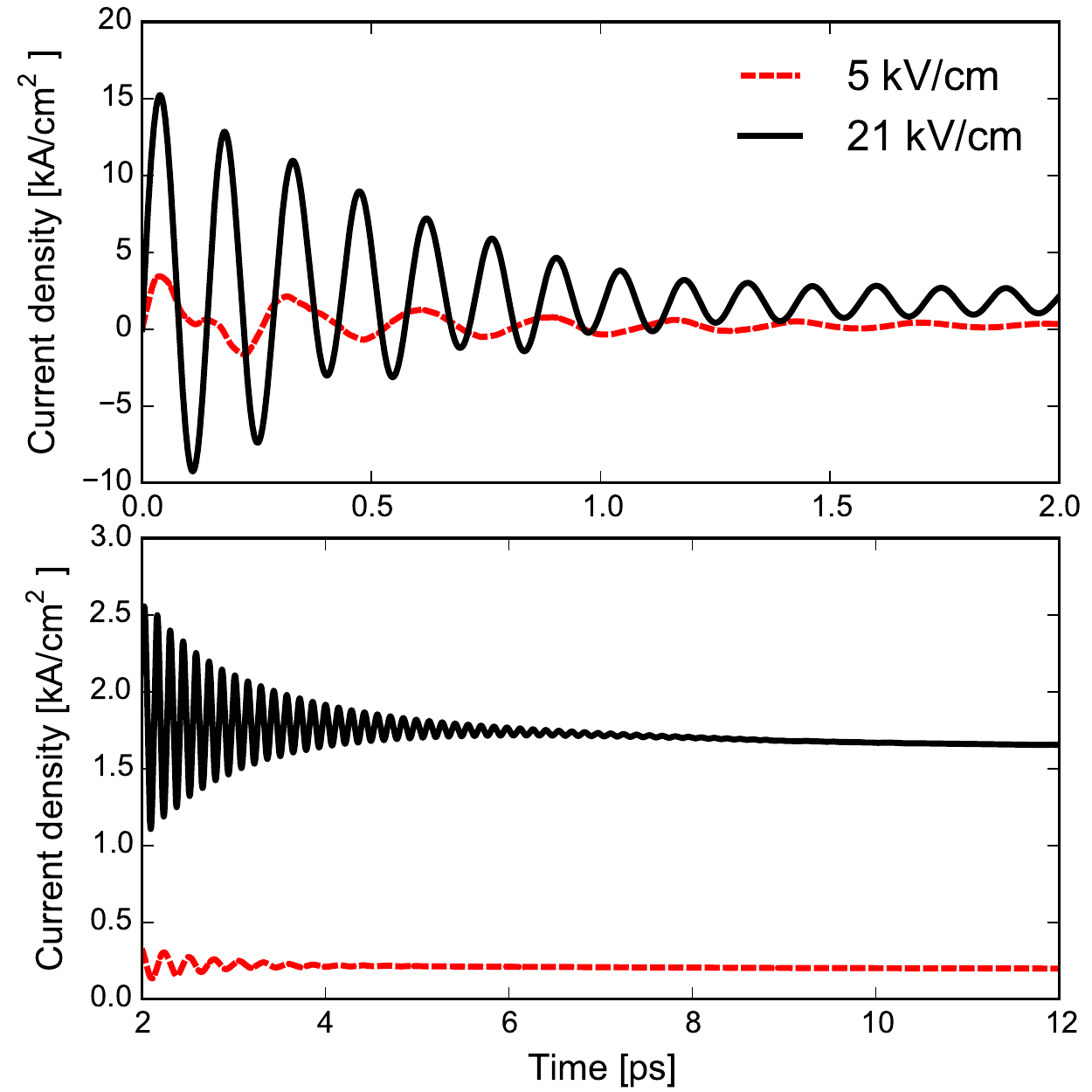}
\caption{Current density vs time for two values of electric fields. The upper panel shows the first two picoseconds and the lower the next 10 picoseconds. Note the different ranges on the vertical axis.}
\label{current_vs_time}
\end{figure}

Figure~\ref{current_vs_time} shows the current density vs time at the design electric field ($21$~kV/cm) and at a lower electric field ($5$~kV/cm). The top  panel shows the initial transient (first $2$~ps) and the bottom panel shows the next $10$~ps, which is long enough for the current density to reach a steady state. In the first ~$2$~ps, we observe high-amplitude coherent oscillations in current, with a period of $100$ to $200$~fs. The rapid coherent oscillations decay on a time scale of a few picoseconds, with the high-bias oscillations decaying more slowly. Note that the peak value of current early in the transient can be more than $10$ times higher than the steady-state value. In the bottom panel, we see a slow change in current, which is related to the redistribution of electrons within subbands, as well as between different subbands.

\begin{figure}
\centering
\includegraphics[width=3.3in]{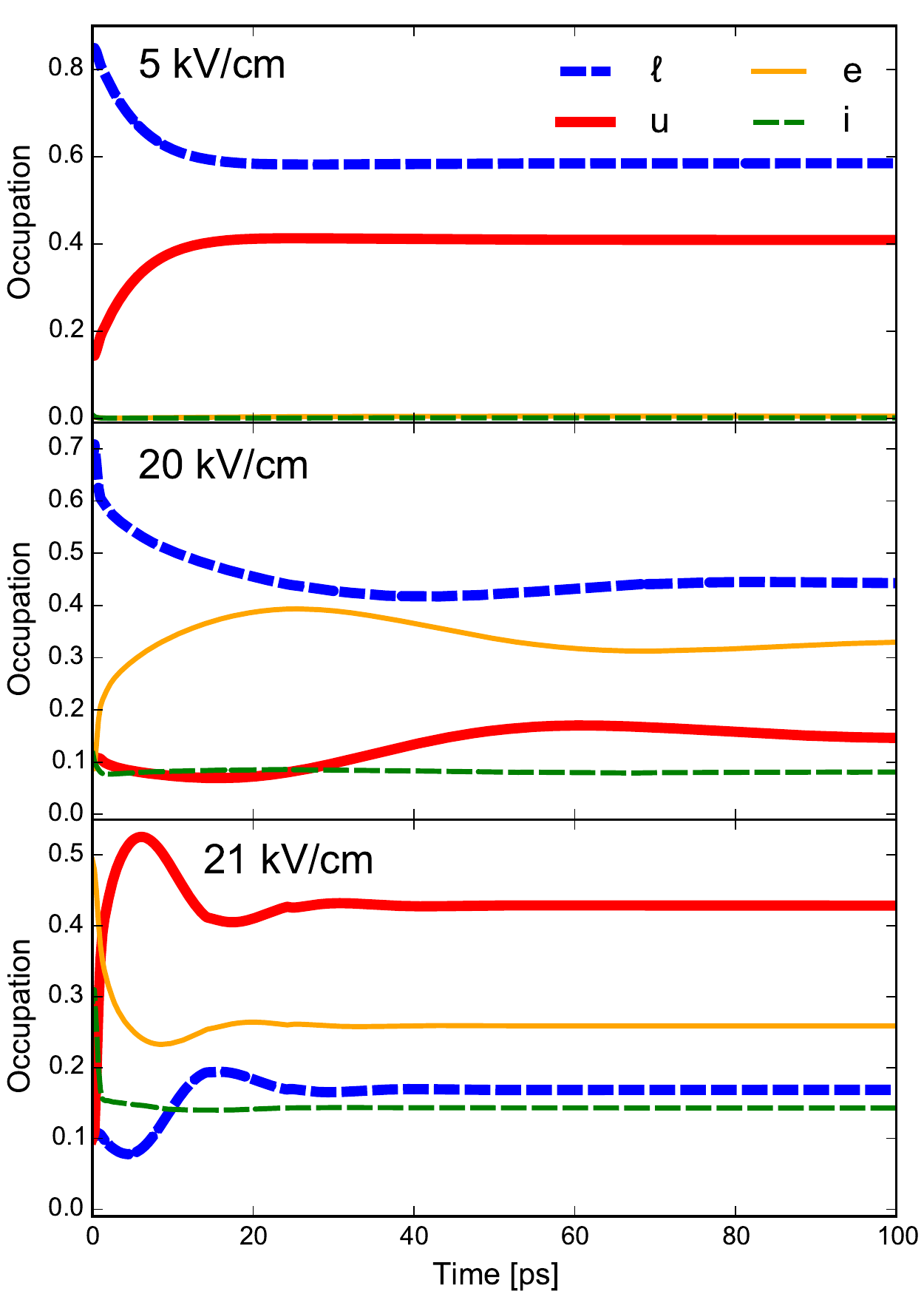}
\caption{Occupation vs time for the lower lasing, upper lasing, injector, and extractor states. Normalization is chosen such that all occupations add up to one. Note the longer time scale compared with the current density in Fig.~\ref{current_vs_time}.}
\label{occupations_vs_time}
\end{figure}

Figure~\ref{occupations_vs_time} shows the time evolution of occupations for the same values of bias as in Fig~\ref{current_vs_time}, in addition to results slightly below the design electric field. Occupations are a very important quantity because the optical gain of the device is directly proportional to the population difference of the upper and lower lasing level ($\rho_{\mathrm{uu}}-\rho_{\ell\ell}$). The time evolution of the occupations tells us how long it takes the device to reach its steady state lasing capability. In Fig~\ref{current_vs_time}, we can see that the occupations take a much longer time to reach steady state ($20$-$100$~ps) than the current density, and the time needed to reach a steady state is not a monotonically increasing function of the electric field: the $20$-kV/cm results take more than twice as long to reach a steady state than the $21$-kV/cm results. In Fig.~\ref{occupations_vs_time}, we see that a population inversion of $\rho_{\mathrm{uu}}-\rho_{\ell\ell}=0.26$ is obtained at the design electric field, while lower-field results show no population inversion.

\begin{figure}
\centering
\includegraphics[width=3.3in]{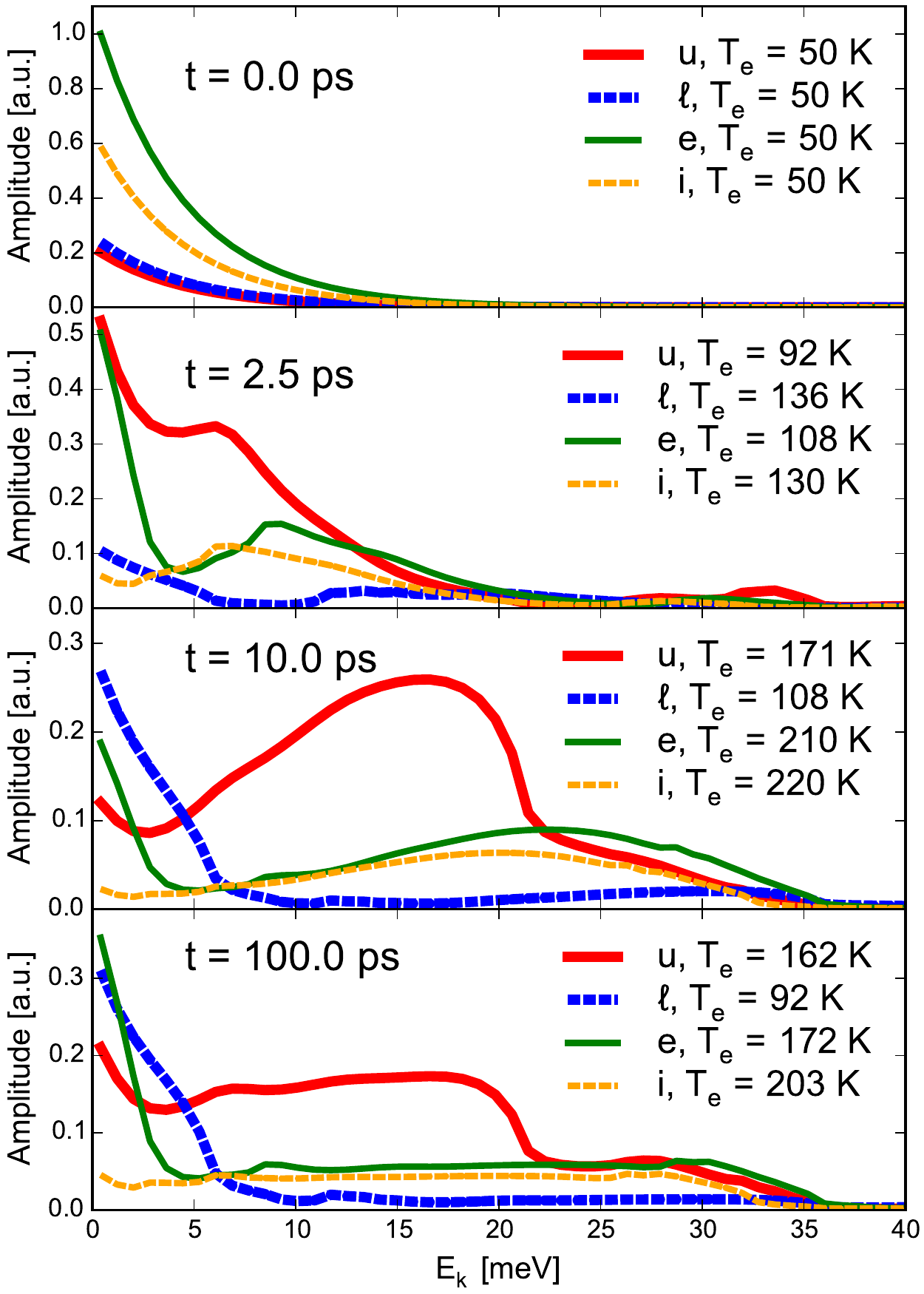}
\caption{In-plane energy distribution for all eigenstates shown in Fig.~\ref{bandstructure}, except for the high energy state (h). Results are shown for four values of time, starting in thermal equilibrium ($t=0$~ps) and ending in the steady state~($t=100$~ps). Also shown are the corresponding electron temperatures, calculated from $T_e=\left< E_k \right>/k_B$.}
\label{inplane_vs_time}
\end{figure}

Figure~\ref{inplane_vs_time} shows the time evolution of the in-plane energy distribution for all the subbands shown in Fig.~\ref{bandstructure}. Also shown is the equivalent electron temperature of each subband calculated using $\left< E_k \right>=k_BT_e$. The top panel showns the initial (thermal equilibrium) state, where all subbands have a Maxwellian distribution with the extractor having the highest occupation. At time $t=2.5$~ps, the in-plane distribution has heated considerably for all subbands, with the lower lasing level being hottest at $T_e=136$~K. At $t=10$~ps, the lower lasing level has cooled while the other states have heated, with the injector state having the highest temperature of $220$~K. At $t=100$~ps, the system has reached a steady state, where the lower lasing level is considerably cooler ($92$~K) than other states, with the injector state being hottest with a temperature of $203$~K. A noticable feature in Fig.~\ref{inplane_vs_time} is the big difference in temperature of the different subbands with a temperature difference of $110$~K between the injector and lower lasing level. In addition to having very different temperatures, the in-plane energy distributions are very different from heated Maxwellian distribution and different subbands have very different in-plane distributions. A weighted average (using occupations as weights) of the steady state electron temperatures is $159$~K, which is $109$~K higher than the lattice temperature.

\section{Conclusion}
\label{sec:conclusion}
We derived a Markovian master equation \eqref{eq:MMEPC} for the single-electron density matrix, including off-diagonal matrix elements (coherences) as well as in-plane dynamics. The MME conserves the positivity of the density matrix, and accounts for scattering of electrons with phonons and impurities. We applied the MME to simulate electron transport in a THz QCL. Close to lasing (around the design electric field), our results for current density are in good agreement with both experiment and theoretical results based on NEGF. The differences between NEGF and density matrix at low fields are small and can be attributed to the omission of interface roughness scattering in our simulation and the effects of collisional broadening.

We have shown that the magnitude of the off-diagonal density matrix elements can be a significant fraction of the largest diagonal element. With the device biased for lasing, the greatest coherence was between the injector and extractor levels, with a magnitude of $21\%$ of the largest diagonal element (the upper lasing level). This results demonstrates the need to include coherences when describing QCLs in the THz range.

We have found that significant electron heating takes place at the design electric field, with in-plane distributions deviating far from a heated Maxwellian distribution. The electron temperature was found to  vary strongly between subbands, with an average subband temperature about $109$~K hotter than the lattice temperature of $50$~K. This result demonstrates the need to treat in-plane dynamics in detail.

Time-resolved results showed that, early in the transient, current density exhibits high-amplitude coherent oscillations with a period of $100$-$200$~fs, decaying to a constant value on a time scale of $3$-$10$ picoseconds. The amplitude of current oscillations could be over $10$ times larger than the steady-state current. Occupations of subbands and in-plane energy distributions took considerably longer ($20$-$100$~ps) than current to reach the steady state.

Solving the MME for the density matrix is a numerically efficient approach to time-dependent quantum transport in nanostructures far from equilibrium.

\begin{acknowledgements}
The authors gratefully acknowledge support by the U.S. Department of Energy, Basic Energy Sciences, Division of Materials Sciences and Engineering, Physical Behavior of Materials Program, Award No. {DE-SC0008712}. The work was performed using the resources of the UW-Madison Center for High Throughput Computing (CHTC).
\end{acknowledgements}


\newpage

\appendix

\section{Calculation of $\mathcal G$ terms}
\label{app:Gterms}

This appendix is devoted to explicit calculation of $\mathcal G$. This task involves the evaluation of Eq.~\eqref{eq:G_def1} for different scattering mechanisms, which is repeated here for convenience
\begin{align}
  \label{eq:G_def2}
  \mathcal G_g^\mathrm{em}(E_k,Q_z,E_q)\equiv \frac{1}{2\pi} \int_{0}^{2\pi}d\theta\mathcal W_g^\mathrm{em}(\ve Q,E_k) \ .
\end{align}
The matrix element $\mathcal W_g^\mathrm{em}(\ve Q,E_k)$ can always be written in terms of $E_q=\hbar^2q^2/2m^*$, $E_k=\hbar^2k^2/2m^*$, $E_z=\hbar^2Q_z^2/2m^*$ and the angle $\theta$ between $\ve k$ and $\ve q$. Note that this definition of $E_z$ is only to make expressions more compact and readable, the actual energy in the $z$-direction is contained in the $\Delta_{nm}$ terms. Derivations of the various phonon matrix elements $\mathcal W_g^\mathrm{em}(\ve Q)$ used in this section can be found in Refs.~\cite{ferry2013,jacoboni1989}.

For the case of longitudinal acoustic (LA) phonons, we employ the equipartition approximation and get
\begin{align}
  \mathcal W_\mathrm{LA}^\mathrm{em}(\ve Q)\simeq  \frac{D_{ac}^2k_BT_L}{2m^*v_s^2}=\beta_{LA} \ ,
\end{align}
where $D_{ac}$ is the deformation potential for acoustic phonons and $v_s$ is the sound velocity in the material. In this case, the $\theta$ integration in Eq.~\eqref{eq:G_def2} gives $\mathcal G_\mathrm{LA}^\mathrm{em}=\beta_{LA}$. Since acoustic phonons are treated elastically, the emission and absorption terms are identical.

As with the acoustic phonons, the nonpolar optical phonon scattering is isotropic, so the phonon matrix element is constant  $\mathcal W_\mathrm{op}^\mathrm{em}(\ve Q)=(N_\mathrm{op}+1)\beta_\mathrm{op}$. The angular integration in Eq.~\eqref{eq:G_def2} gives $\mathcal G_\mathrm{op}^\mathrm{em}=(N_\mathrm{op}+1)\beta_\mathrm{op}$.

The phonon matrix element for electron scattering with polar optical phonons,  with screening included, is given by
\begin{subequations}
\begin{align}
  \mathcal W_\mathrm{pop}^\mathrm{em}(\ve Q)=(N_\mathrm{pop}+1)\beta_\mathrm{pop} \frac{Q^2}{(Q^2+Q_D^2)^2} \ ,
\end{align}
where $Q_D$ is the the Debye wave vector defined by $Q_D^2=ne^2/(\eps k_B T)$ and
\begin{align}
  \beta_\mathrm{pop}=\frac{e^2E_\mathrm{pop}}{2\eps_0} \left( \frac{1}{\eps_r^\infty} - \frac{1}{\eps_r} \right) \ ,
\end{align}
\end{subequations}
where $\eps_r^\infty$ and $\eps_r$ are the high-frequency and low-frequency relative permittivities of the material, respectively and $n$ is the average electron density. The effects of screening are quite small at the electron densities considered in this work, however the $1/Q^2$ singularity poses problems in numerical calculations due to the high strength of the POP interaction. These problems are avoided by including screening. We can now calculate
\begin{align}
\nonumber
&\frac{\mathcal G_\mathrm{pop}^\mathrm{em}(E_k,E_z,E_q)}{(N_\mathrm{pop}+1)}
  =\frac{\beta_\mathrm{pop}}{2\pi} \int_0^{2\pi} d\theta \frac{ |\ve Q-\ve k|^2 }{( |\ve Q-\ve k|^2+Q_D^2 )^2} \\ \nonumber
&= \frac{\beta_\mathrm{pop}}{2\pi}\int_0^{2\pi} d\theta \frac{Q_z^2+q^2+k^2-2qk\cos(\theta)}{( q_z^2+q^2+k^2-2kq\cos(\theta) +Q_D^2 )^2} \\ \nonumber
&= \frac{\beta_\mathrm{pop}}{Q_z^2+q^2+k^2} \left[  1+\frac{Q_D^2}{Q_z^2+q^2+k^2} - \frac{4k^2q^2}{(Q_z^2+q^2+k^2)^2 } \right] \times \\ \nonumber
&\left[ \left(1+\frac{Q_D^2}{Q_z^2+q^2+k^2}\right)^2 -\frac{4k^2q^2}{(Q_z^2+q^2+k^2)^2}  \right]^{-\frac32} \\ \nonumber
&= \frac{\hbar^2}{2m^*}\frac{\beta_\mathrm{pop}}{E_z+E_k+E_q} \times \\  \nonumber
&\left[  1+\frac{E_D}{E_z+E_q+E_k} - \frac{4E_kE_q}{(E_z+E_q+E_k)^2 } \right] \times \\
&\left[ \left(1+\frac{E_D}{E_z+E_q+E_k}\right)^2 -\frac{4E_kE_q}{(E_z+E_q+E_k)^2}  \right]^{-\frac32} \ ,
\end{align}
where $E_D=\hbar^2Q_D^2/2m^*$ is the Debye energy.

The matrix element for ionized impurities is given by
\begin{subequations}
\begin{align}
  \mathcal W_\mathrm{ii}^\mathrm{em}(\ve Q)=\frac{\beta_{ii}}{|\ve Q|^4}
\end{align}
with
\begin{align}
  \beta_{ii}=\frac{N_IZ^2e^4}{2\eps_r^2\eps_0^2}\ ,
\end{align}
\end{subequations}
where $N_I$ is the impurity density, and $Z$ is the number of unit charges per impurity. This matrix element gives
\begin{align}
\nonumber
&\mathcal G_\mathrm{ii}^\mathrm{em}(E_k,E_z,E_q)=\frac{\beta_{ii}}{2\pi} \int_0^{2\pi} d \theta \frac{1}{(Q_z^2+q^2+k^2-2qk\cos(\theta))^2}\\ \nonumber
&= \beta_{ii} \frac{Q_z^2+q^2+k^2}{[ (Q_z^2+q^2+k^2)^2-4q^2k^2 ]^{3/2}} \\
&=\beta_{ii}\frac{\hbar^4}{4m^2}\frac{E_z+E_k+E_q}{[(E_z+E_k+E_q)^2-4E_kE_q]^{3/2}} \ .
\end{align}
Since ionized-impurity scattering is elastic, the absorption term is identical to the emission term.

\end{document}